\documentclass[prl,twocolumn]{revtex4}
\usepackage{graphicx}
\usepackage{amsmath,amssymb,amsthm,amsbsy}
\usepackage{latexsym}
\usepackage{amsfonts}
\usepackage{amssymb}

\begin{document}

\title{Simulation of Coherent Synchrotron Radiation Emission from
Rotating Relativistic Electron Layers}

\author{Bjoern S. Schmekel}
\affiliation{Department of Physics,
Cornell University, Ithaca, New York 14853}
\email{bss28@cornell.edu}

\begin{abstract}
The electromagnetic radiation of rotating relativistic electron layers
is studied numerically using particle-in-cell simulation. 
This system is general enough to be applied to problems involving
the coherent synchrotron radiation (CSR) instability in bunch
compressors and possibly in radio pulsars where the CSR instability
may be used to explain the high brightness temperature and the
observed spectra. The results of the simulation confirm all relevant 
scaling properties predicted by theoretical models and suggest that
their range of validity is bigger than conservative estimates indicate.
This is important for radio pulsars where the allowed parameter range
is unknown.
\end{abstract}

\maketitle

Extremely high brightness temperatures
encountered for example in the radio emission of 
pulsars cannot be explained by an incoherent 
radiation mechanism for which
the radiated power is proportional to the number
of radiating charges. However, if all outgoing waves
interfere constructively the total power scales as
the square of that number.
In future particle accelerators the wavelength
of the emitted radiation can easily reach the
length of the accelerated bunches such that
bunches can modulate themselves by means of
their own radiation. The possible energy loss
caused by CSR would be undesirable for the
operation of such accelerators.

The aim of the present work is to test the results
obtained from the model developed by Schmekel, Lovelace
and Wasserman \cite{schmekel2004} numerically and probe its
range of validity. The model predicts growth
rates, saturation amplitudes and the emitted power
for an initial perturbation of a rotating relativistic
cylinder of charged particles in an external magnetic
field. The linear instability is closely related to the instability
found by Goldreich and Keeley \cite{GoldreichKeeley1971}.
More recently CSR has been investigated in the particle accelerator
physics community, e.g. \cite{Stupakov2002,Heifets2002,Byrd2003,Venturini2005}.

In \cite{schmekel2004} the distribution function is chosen such that the azimuthal
canonical angular momentum is fixed and the energy drops off exponentially.
For the time being all fields and distribution
functions are assumed to be uniform in the z-direction.
The equilibrium can be parametrized by the radius $r_0$ of the
cylinder, the relative energy spread $\delta E / E \equiv v_{th}^2$ of the particles,
the Lorentz factor $\gamma$ and the dimensionless field reversal parameter
\begin{eqnarray}
\zeta \equiv \frac{4 \pi e n_0 v_{\phi} r_0 v_{th} \sqrt{\pi/2}}{B_z^{ext}} ,
\end{eqnarray}
where $n_0$ is the central number density, $v_{\phi}$ the azimuthal velocity, 
$B_z^{ext}$ is the external magnetic field in the $z$-direction in cgs units and $c=1$.
Considering all perturbed quantities to have the dependence $\exp \left ( im \phi - i \omega t + k_r r \right )$
one obtains for $m^{2/3} v_{th} \ll 1$, $k_r v_{th} r_0 \ll 1$ and $|\Delta \tilde \omega | \ll \gamma^{-2}$ 
the dispersion relation
\begin{eqnarray}
1=-i \pi \zeta J_m(\omega r_0) H_m(\omega r_0) \gamma^{-2} (\Delta \tilde \omega)^{-2} F_0
\label{dispF0}
\end{eqnarray}
and the growth rate
\begin{equation}
\Im(\omega) \simeq {1.083\zeta^{1/2}m^{2/3}\dot\phi\over\gamma} \sqrt{F_0} ,
\label{analyticm1}
\end{equation}
where $\Delta \tilde \omega = (\omega - m \dot \phi)/(m \dot \phi)$,
$H_m = J_m + i Y_m$ with the Bessel functions $J_m$ and $Y_m$,  
$\dot \phi$ is the angular velocity, the dot denotes the time 
derivative, and
\begin{eqnarray}
F_0 \equiv \frac{1}{2\pi} \int _{-\pi }^{\pi }
e^{-\frac{m^2 v_{th}^2}{2} \left ( 1 - \sin 
\theta \right )^2} d\theta .
\label{Fnnum}
\end{eqnarray}
The radiated power is given by
\begin{eqnarray}
P_m \approx 3.71 \times 10^{14}
   \gamma^6 m^{-3} \frac{L}{r_0}
\left ( \frac{\Im(\omega)}{\dot \phi} \right )^4
\frac{\rm erg}{\rm s}~.
\label{power}
\end{eqnarray}
Here the length of the cylindrical layer is denoted by $L$.
The quoted expression was derived by approximating Bessel functions with Airy functions.
It was shown that this approximation is valid under the conditions
$m \gg 1$ and $m \gg \zeta^{3/2}\gamma^3 \equiv m_1$.
The used dispersion relation was derived under the assumption $|\Delta \tilde \omega | \ll \gamma^{-2}$
(which is approximately equivalent to $m \gg m_1$),
but this condition may be too strict. There is evidence that this is indeed the case and one 
objective of this article is to confirm this. Especially for the study of radio pulsars
bigger growth rates may be necessary to explain the observed brightness temperature.

For the simulation the software package OOPIC \cite{verboncoeur} was used.
OOPIC is a relativistic two-dimensional particle-in-cell
code which supports both plain $(x,y)$-geometries and cylindrical
$(r,z)$-geometries.
Since the interesting dynamics takes places in the azimuthal
direction one can only simulate a thin ring (instead of a cylinder)
in the $(x,y)$-mode. Loading the initial circular particle distribution
in the $(x,y)$-mode required modifying the source code (files load.cpp, diagn.cpp
and c\_utils.c) to allow the program to handle circular particle distributions.
Some minor modifications were necessary in order to compile XOOPIC-2.5.1 with 
gcc 3.2.2 and the compiler compiler bison 1.28 under SunOS 5.9. 
The built-in function parser was extended to support elliptic integrals.

Since a thin ring of particles is simulated instead of a cylinder thereof all fields and charges
were divided by the length $L$ of the cylinder whereas the electron
mass needs to be divided by $L^2$. $L=10 r_0$ is chosen unless noted otherwise.
The electric and magnetic self-fields for a thin ring
equilibrium differ from what was used in the model.
The fields can be found in \cite{jackson}. It is ensured that OOPIC
uses these self-fields before the perturbation starts to build up.
As it turns out choosing the correct self-fields is not too crucial.
Leaving them out the system will build them up itself. Once the 
self-fields are created the system shows no difference in behavior.
The absence of the self-fields in the dispersion relation
might help to understand this feature.
As in \cite{schmekel2004} a Gaussian number density profile with RMS
width $v_{th}$ was chosen for the initial distribution. 
$5000$ macro particles were tracked on a grid
with resolution $512 \times 512$ unless noted otherwise.
Once an energy for a particle has been chosen it is placed at the equilibrium
radius $r_0 = m \gamma c (eB)^{-1}$, i.e. neglecting betatron oscillations
particles on the same orbit have the same energy. This fixes the azimuthal 
component of the canonical angular momentum. The system can
pick up transverse motion quickly.
The grid represents a rectangular region $40 {\rm m} \times 40 {\rm m}$ big where 
the ring with radius $r_0=10 {\rm m}$ is centered.

\begin{figure}
\includegraphics[width=3.5in]{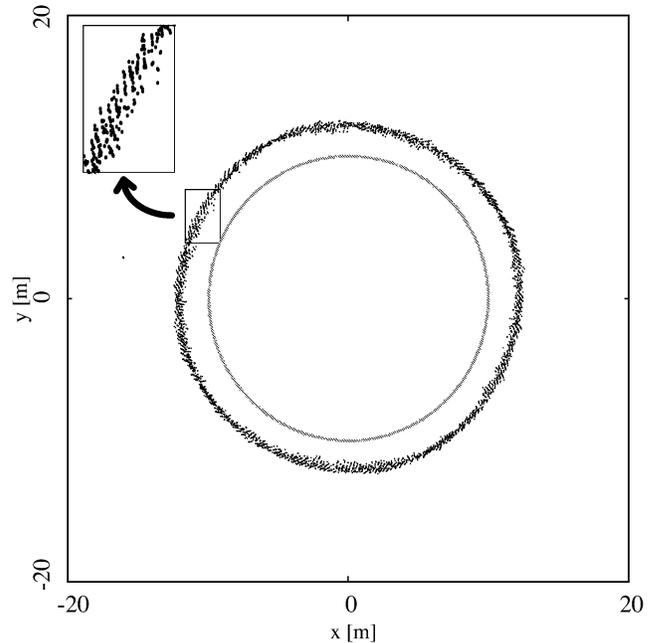}
\caption{Initial particle distribution (gray)
and the same distribution after 23ns have elapsed.
Parameters: $\zeta=0.010$, $\gamma=30$ and $v_{th}=0.002$}
\label{ely71_p23e8}
\end{figure}

\begin{figure}
\includegraphics[width=3.5in]{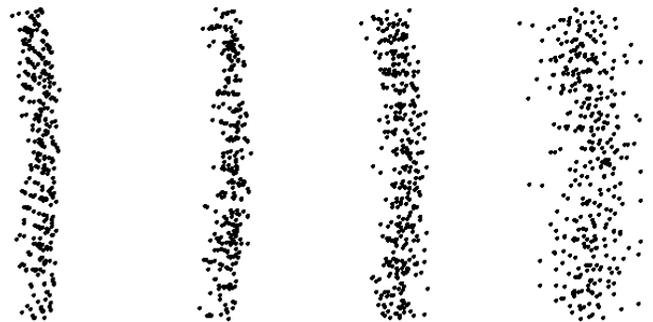}
\caption{Particle distribution ($\gamma=30$, $\zeta=0.01$) after 23ns
for $v_{th}=0.002$, $v_{th}=0.008$, $v_{th}=0.015$
and $v_{th}=0.033$ (from left to right). The four figures depict an
approximately 6.5m long segment.} 
\label{bunchvth}
\end{figure}

In Fig. \ref{ely71_p23e8} the initial particle distribution (gray) and the particle distribution 
after $23ns$ are shown. The parameters are $\zeta=0.010$, $\gamma=30$, and $v_{th}=0.002$.
Qualitatively, a bunching of the particle distribution can be observed.
An enlargement of a small section of Fig. \ref{ely71_p23e8} is also shown in the same figure.
In Fig. \ref{bunchvth}, the bunching is shown for successively higher energy spreads.
With increasing energy spread the bunches become fuzzier and the clean gaps between bunches that
can be observed for small energy spreads are populated with ``stray particles''. This suggests
that it may be harder to achieve complete coherence for larger values of $v_{th}$. The decoherence
due to the non-zero width of the particle beam is investigated quantitatively later in the paper. 
These qualitative features are independent of $\gamma$.

\begin{figure}
\includegraphics[width=3.5in]{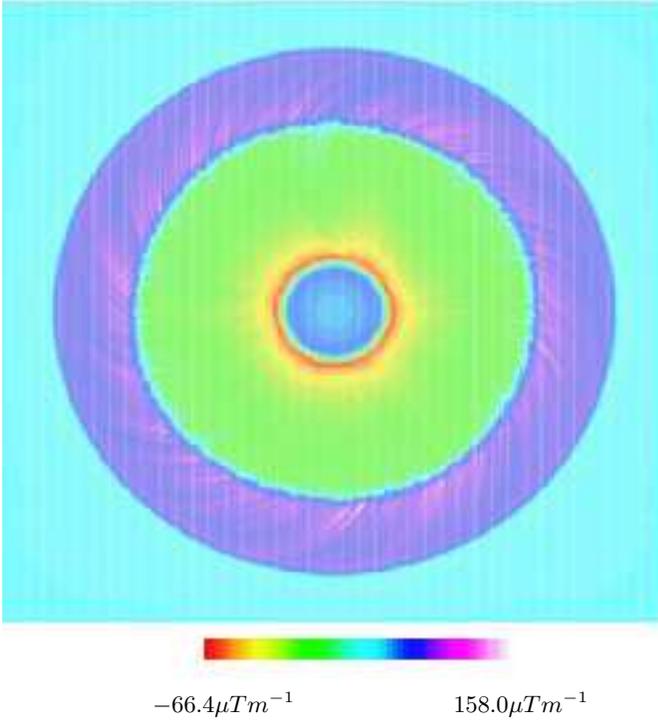}
\setlength{\unitlength}{1mm}
\flushleft\begin{picture}(0,0)(0,0)
\put(20,3){$-66.4 \mu T m^{-1}$}
\put(60,3){$158.0 \mu T m^{-1}$}
\end{picture} 
\caption{$z$-component of the magnetic field (self-field
plus perturbation without external magnetic field) after
23ns for $\zeta=0.010$, $\gamma=30$ and $v_{th}=0.025$.
The size of the area depicted is 40m $\times$ 40m.}
\label{ely84_c23e8}
\end{figure}

Also note that during the evolution of the circular charge distribution both the radius and the width
of the ring increase slightly. The former is due to a.) particles losing energy
and b.) the perturbed magnetic field changing significantly. It tends to decrease for
small energy spreads and increase for larger energy spreads.
Starting with a larger radius the radius increases even further, i.e. this is not a relaxation
from a ``false'' to a true equilibrium.
Since the non-zero mesh size imposes an upper limit on the azimuthal mode number $m$ which
can become unstable, it is expected that the distance between bunches
decreases as the resolution increases. This is indeed the case.
For larger energy spreads the bunching of the distribution becomes hardly visible, but it 
still can be observed in the $z$-component of the magnetic field (Fig. \ref{ely84_c23e8}).

The bunches are slightly tilted and may be connected by
a very thin inner ring of particles for sufficiently high beam currents. For these
reasons it is not possible to Fourier transform the charge perturbations in order to
compute the growth rates for each value of $m$. Since the resolutions used were
low the range of $m$ values is restricted. Therefore, only the radiated power is computed which
can be obtained easily.
%
%
%
%
Estimates of the growth rate are two orders of magnitude higher than what would be expected.
A possible explanation is that the ratio between the saturation amplitude and
the electric self field 
\begin{eqnarray}
\bigg | \frac{\delta E_{sat}}{E_{self}} \bigg | =
\frac{1}{m \zeta} \left({\Im(\omega)(m) \over \dot{\phi}}\right)^2
\end{eqnarray}
is typically in the order of $10^{-3}$ for the given sample cases
which is rather small. The initial perturbations due to discreteness,
numerical noise etc. are usually in the same order of magnitude.
Therefore, one cannot expect to see the regime covered by the linearized
Vlasov equation. This is another reason for focusing entirely on
the emitted power.

%
\begin{figure}
\includegraphics[width=3.5in]{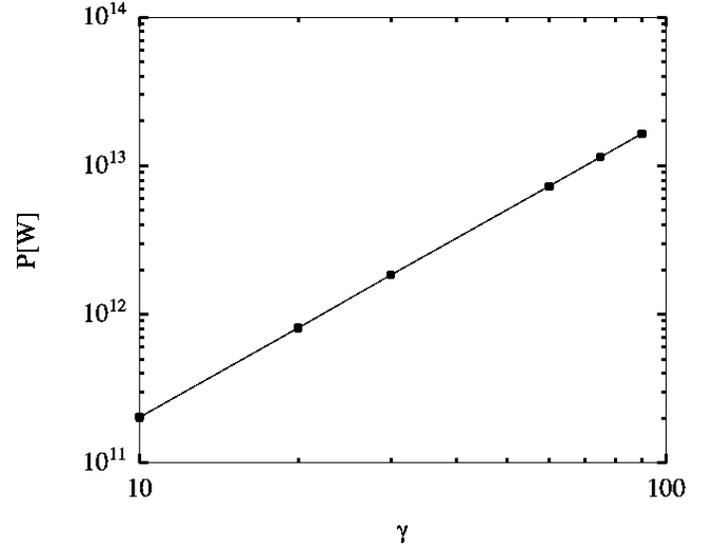}
\caption{Loss of kinetic energy in W vs.
$\gamma$ for $\zeta=0.01$ and $v_{th}=0.002$.}
\label{gamma}
\end{figure}
%
\begin{figure}
\includegraphics[width=3.5in]{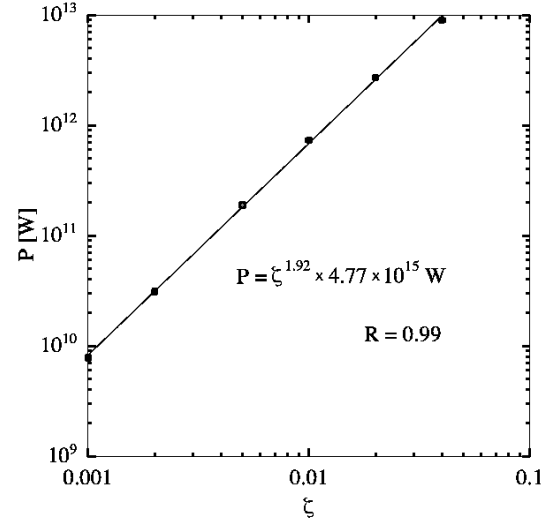}
\caption{Loss of kinetic energy in W vs.
$\zeta$ for $\gamma=30$ and $v_{th}=0.025$.
The solid line shows the best fit.}
\label{powerfit}
\end{figure}
%

In Fig. \ref{gamma} and Fig. \ref{powerfit} the radiated power determined by measuring
the kinetic energy loss of the electron cloud after approximately 2.36 ns
is plotted as a function of $\gamma$ and $\zeta$, respectively. After 0.24 ns the
perturbations have saturated and the emitted power is fairly constant.
A quadratic dependence can be established, i.e. the first two relevant scalings 
expected from the analytical model are recovered.

The simulation has been repeated at lower (256 $\times$ 256) and higher (1024 $\times$ 1024) resolution.
No significant effect could be observed. This is consistent
with the model which predicts that most power is emitted by modes with 
low $m$. Also, decreasing the stepsize $dt$ to 2.5 ps and increasing the
number of macro particles to 50000 has a negligible effect.

\begin{figure}
\includegraphics[width=3.5in]{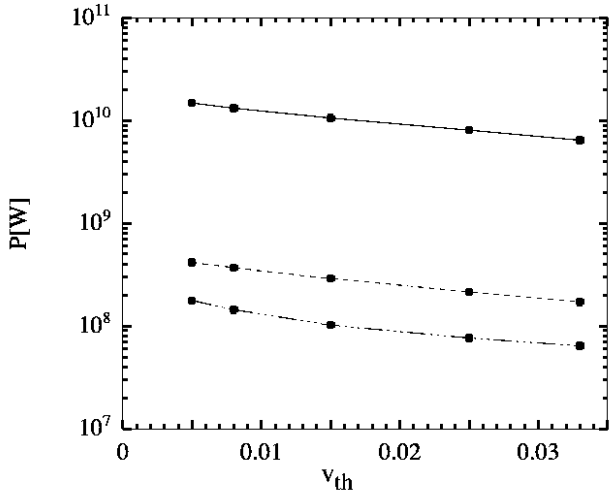}
\caption{Total power radiated as obtained from OOPIC for the parameters
$\zeta=0.01$, $\gamma=30$ (solid) and $\zeta = 0.005$, $\gamma = 10$ (dashed).
The dash-dotted line is proportional to $\sum_{m=1}^{\infty} P_m$.}
\label{compvth}
\end{figure}

Finally, the effect of the energy spread is investigated.
With increasing $v_{th}$ the power decreases
which is due to the decoherence described by the factor $F_0$
defined in Eq. \ref{Fnnum}. The results are plotted in
Fig. \ref{compvth} for the parameters 
$\zeta=0.01$, $\gamma=30$ and $\zeta = 0.005$, $\gamma = 10$, respectively, and $L=r_0$.
In the former case $m_1$ is 27 and in the latter case it is 0.4.
Eq. \ref{power} becomes 
\begin{eqnarray}
P \approx 3.71 \times 10^{14} \frac{\rm erg}{\rm s}  \frac{L}{r_0} \gamma^2 
\sum_m m \left [ i \pi J_m(\omega r_0) H_m(\omega r_0) \right ]^2
\end{eqnarray}
Despite $m_1$ being much larger than 1 for the first set of parameters the 
slopes in Fig. \ref{compvth} match exactly only if the summation starts
at $m=1$.  This suggests that modes with $m < m_1$ do radiate and can
be described by the same dispersion relation. Since the power scales as $m^{-5/3}$   
these modes may actually be very important for computing the total energy loss.
Note that while the simulation suggests $P \propto L^2$ Eq. \ref{power} (which was
derived under the assumption $L \gtrsim r_0$) gives
$P \propto L$. A 2D simulation cannot explain how the radiation from different
axial positions on the cylinder interacts. In the thin ring case doubling $L$
doubles the number of particles $N$ and therefore quadruples $P$.
Fortunately, as can be seen in the derivation of Eq. \ref{power} in \cite{schmekel2004} 
the $\zeta$, $v_{th}$ and $\gamma$ dependent part of $P$ is independent of $L$.
In Fig. \ref{compvth} the overall factor matches if $r_0 = 100L$ whereas the growth
rate for a perturbation of a cylinder and a thin ring coincide for $r_0 = L$ \cite{schmekel2004}.  
Also note that Fig. \ref{ely71_p23e8} suggests $k_r v_{th} r_0 \sim 1$, whereas Eq. \ref{analyticm1} was derived
under the assumption $k_r v_{th} r_0 \ll 1$.

%
%

The particle in cell code OOPIC was used to simulate the evolution
of density perturbations in a thin ring of charged particles which
move in relativistic almost circular motion in an external magnetic field.
The results were compared with the model in \cite{schmekel2004}.
Comparisons of the simulation with the model shows approximate agreement 
with the main predicted scaling relations.
In particular the bunching effect
could be observed very clearly and the emitted power is proportional to
the square on the number density which implies coherent radiation.
The dependence on the energy spread can be recovered exactly assuming
all modes contribute to the observed energy loss suggesting that
the model may apply even if $m < m_1$.  

\acknowledgments
I thank Richard V.E. Lovelace, Georg H. Hoffstaetter,
Ira M. Wasserman and Joseph T. Rogers for valuable discussions and John P. Verboncoeur
for his help with OOPIC. I also thank the referee for pointing out one mistake.
This research was supported by the Stewardship Sciences Academic Alliances program of the National Nuclear
Security Administration under US Department of Energy Cooperative agreement DE-FC03-02NA00057.

\end{document}